\documentclass[a4paper,showpacs,twocolumn,floatfix,prb,citeautoscript,superscriptaddress]{revtex4}

\usepackage[dvips]{graphicx}
\usepackage{dcolumn}% Align table columns on decimal point
\usepackage{bm}% bold math
\usepackage{amsmath}
\usepackage{amssymb}
\usepackage{times}

\graphicspath{{.}{./EPS/}}

\begin{document}

\pacs{74.45.+c,71.70.Di,73.20.-r,73.40.-c}
% 71.70.Di (Landau levels), 73.20.-r (Surface and interface electron state), 73.40.-c (Electronic transport in interface structures), 74.45.+c (Proximity effects; Andreev effect; SN and SNS junctions)
%\preprint{}

\title{Andreev reflection and cyclotron motion at superconductor -- 
normal-metal interfaces} 

\author{F. Giazotto}
%\email{giazotto@sns.it}
\affiliation{NEST-INFM and Scuola Normale Superiore, Piazza dei Cavalieri 7,
I-56126 Pisa, Italy}

\author{M. Governale}
\affiliation{NEST-INFM and Scuola Normale Superiore, Piazza dei Cavalieri 7,
I-56126 Pisa, Italy}

\author{U. Z\"ulicke} 
\affiliation{Institute of Fundamental Sciences and MacDiarmid Institute for
Advanced Materials and Nanotechnology, Massey University, Private Bag
11~222, Palmerston North, New Zealand}

\author{F. Beltram}
\affiliation{NEST-INFM and Scuola Normale Superiore, Piazza dei Cavalieri 7,
I-56126 Pisa, Italy}

\date{\today}

\begin{abstract}

We investigate Andreev reflection at the interface between a superconductor
and a two--dimensional electron system (2DES) in an external magnetic field
such that cyclotron motion is important in the latter. A finite Zeeman splitting in
the 2DES and the presence of diamagnetic screening currents in the
superconductor are incorporated into a microscopic theory of Andreev edge
states, which is based on the Bogoliubov--de~Gennes formalism. The
Andreev--reflection contribution to the interface conductance is calculated.
The effect of Zeeman splitting is most visible as a double--step feature in the
conductance through clean interfaces. Due to a screening current, 
conductance steps are shifted to larger filling factors and the formation of
Andreev edge states is suppressed below a critical filling factor.

\end{abstract} 

\maketitle

\section{Introduction}

The trend toward radical miniaturization of electronic devices has spurred
efforts aimed at understanding the effect of quantum coherence on transport.
Current research is investigating a variety of such {\em mesoscopic\/}
systems~\cite{leshouches95,mesoeltrans,imrybook}. Recently, the subject of
{\em mesoscopic superconductivity}~\cite{hekkingreview,beenreviewsh,
lambertreview} has attracted considerable interest, partly due to the
possibility to perform experiments in hybrid structures consisting of 
superconductors and 
semiconductors where phase-coherence lengths exceed device 
dimensions~\cite{vanweesreviewsh}. Many effects observed in these systems
ulimately stem from {\em Andreev reflection\/}~\cite{andreev1,andreev2}.
Speaking in simple terms, Andreev reflection occurs at the interface between
a superconductor~(S) and a normal metal~(N) when an electron with energy
within the superconducting gap is incident on the interface from the N side.
While no states are available in S for the incoming electron, a finite amplitude
for pairing with some suitable electron from N is induced by the proximity of
S, which allows both electrons to enter S as a Cooper pair. If that happens, N 
is left with a hole excitation that has the sign of its group velocity, charge, and
mass opposite to that of the electron. As incoming electron and
Andreev-reflected hole have a definite phase relation, bound states are
formed in multi-interface geometries such as S--N--S~\cite{kulik} or
S--N--I~\cite{adz:jetp:80} systems, giving rise to discrete energy levels within
the superconducting gap.

An external magnetic field $\vec B=\vec\nabla\times\vec A$ affects transport
properties of mesoscopic systems at least in two ways. In systems with finite
Zeeman splitting (given by $g\mu_{\text{B}} B$, $g$ being the gyromagnetic factor and $\mu_B$ the Bohr magneton), 
spin degeneracy is lifted,
which turns out to suppress Andreev reflection~\cite{been:prl:95,soul:sci:98, 
buhr:prl:98}. In addition, the vector potential $\vec A$ explicitly enters the
single--electron Hamiltonian,
\begin{equation}\label{magham}
H_{\text{0}} = \frac{1}{2 m}\left(\vec p - e \vec A \right)^2 + \frac{g}{2}\, 
\mu_{\text{B}} \,\vec\sigma\cdot\vec B \quad ,
\end{equation}
resulting in a host of peculiar quantum-mechanical effects. One of the most
well-known is perhaps the appearance of phase factors which lead to
magnetoconductance oscillations in a variety of mesoscopic 
systems~\cite{imrybook} and affect orbital wave functions even in regions
where no magnetic field is present~\cite{ahabohm}. If the magnetic field is
treated within a quasiclassical approximation which assumes the  relevant
physical length scale of the system to be much smaller than the cyclotron
radius of electrons, the Hamiltonian of Eq.~(\ref{magham}) reduces to
\begin{equation}
H_{\text{0}}^{\text{(qc)}} = \epsilon_{\text{F}} + v_{\text{F}}\,\,\hat p \cdot \left(
\vec p - e \vec A\right) + \frac{g}{2}\, \mu_{\text{B}} \,\vec\sigma\cdot\vec B 
\quad ,
\end{equation}
and modifications of orbital wave functions due to the magnetic field arise
entirely from such phase factors. Here, $\epsilon_{\text{F}}=\hbar^2 
k_{\text{F}}^2/(2 m)$, $v_{\text{F}}=\hbar k_{\text{F}}/m$, and $k_{\text{F}}$
are the Fermi energy, Fermi velocity, and Fermi wavevector, respectively.
Most of the previous studies of Andreev reflection in a magnetic field were 
carried out in the quasiclassical limit~\cite{galaiko1,galaiko2,gogadze,
svid:sjltp:75,adz:jetp:80,adz:ssc:82,belz:prb:98,urs:prb:99,quasiclass}. 

The quasiclassical approximation ceases to be valid, however, when the
cyclotron radius $r_{\text{c}}=l_B^2 k_{\text{F}}$ is smaller than the 
characteristic length scales of the system which are set, e.g., by disorder,
temperature, or size (here, $l_B=\big(\hbar /|e \vec B|\big)^{1/2}$ denotes
the {\em magnetic length\/} which is the fundamental quantum-mechanical
length scale introduced by the magnetic field.) In this limit, another
well-known quantum effect due to the magnetic field becomes important,
which is Landau quantization~\cite{ldl:zphys:30} of electron motion in planes
perpendicular to the field direction. In transport, Landau quantization 
manifests itself, e.g., by Shubnikov~--~de~Haas oscillations and
by the quantum Hall effect~\cite{qhe-sg,qhe-tc}. Recent advances in fabrication
technology make it possible to study Andreev reflection in
superconductor--semiconductor hybrid systems at magnetic fields where
cyclotron motion of quasiparticles can be important in the semiconducting
region~\cite{htak:physb:98,will:prb:99,schap:prb:00,schap:jap:04,eroms04},
motivating a number of related theoretical studies~\cite{ytak:prb:98a,
uz:prl:00,asa1:prb:00,asa:jpsj:00,asa2:prb:00,uz:physb:01,cht:jetp:01,
tka:prb:04,earlycomment}.

Here we present results on the interplay between Andreev reflection and cyclotron
motion at S--N interfaces. We focus in particular on the effect of a finite
Zeeman splitting in the normal region and on the impact of the diamagnetic screening
current in the superconductor~\cite{tkach:prb:05}. %UZ_2: added this reference
To be specific, we choose a
geometry where $\vec B = B\,\hat z$ and the interface is the $y z$ plane. Considering
particle motion in the $x y$ plane from a classical point of view, the interplay between
cyclotron motion and Andreev reflection at an ideal interface results in
electrons and holes alternating in skipping orbits along the
interface~\cite{aro:jetp:78,altskip} in $y$ direction, as shown in
\begin{figure}[b]
\includegraphics[width=3.3in]{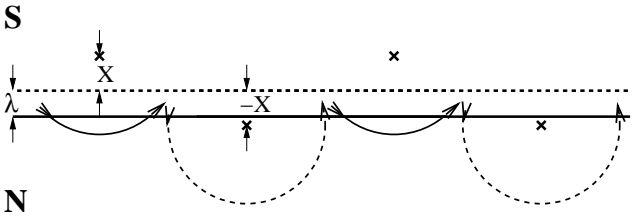}
\caption{Semiclassical picture of Andreev edge states. Electrons and holes
in skipping orbits along a clean S--N interface are successively
Andreev--reflected into each other. In the S region, quasiparticle
excitations exist only on the (extremely short) length scale of the
superconducting coherence length. The presence of a uniform supercurrent
parallel to the interface results in a shift of electron and hole orbits in
the normal region. Andreev reflection couples electron and hole orbits with
guiding--center coordinate $\pm X-\lambda$, respectively. $\lambda$ denotes the magnetic penetration depth. 
\label{fig0}}
\end{figure}
Fig.~\ref{fig0}. In quantum mechanics, however, the guiding-center
coordinates for cyclotron orbits are canonically conjugate operators which
cannot be diagonalized simultaneously~\cite{ahmintrosh} while each of them
separately commutes with $H_0$. In the present problem, where translational 
invariance is broken due to the interface, it is useful to
choose the representation where the guiding-center coordinate $X$ in $x$
direction is a good quantum number, i.e., eigenstates of $H_{\text{0}}$ are
labeled by $X$. Details of our quantum-mechanical description of the S--N
interface in a magnetic field are given in the following section. It is based on
finding solutions of the Bogoliubov~--~de~Gennes (BdG) 
equation~\cite{degennes},
\begin{equation}\label{bdg}
\left(\begin{array}{cc} H_{\text{0}} + U - \epsilon_{\text{F}} &
|\Delta|\,e^{2 i \varphi} \\ |\Delta|\, e^{-2 i \varphi} & -
H_{\text{0}}^* - U + \epsilon_{\text{F}} \end{array}\right)\left(
\begin{array}{c} u \\ v \end{array} \right)= E\, \left(
\begin{array}{c} u \\ v  \end{array}\right)\, ,
\end{equation}
for the hybrid system by matching at the interface appropriate solutions for the S and N regions~\cite{btk}. Scattering at the interface is modeled by an
external potential $U=U(x)=U_{\text{0}}\,\delta(x)$. We allow for different
effective masses and Fermi energies in the S and N regions. For our study,
we adopt the model where the relevant quasiparticle excitations in the S
region are those of a conventional BCS s-wave superconductor with a
uniform superflow in the $y$ direction parallel to the interface due to the
presence of diamagnetic screening currents. This approach is strictly valid
only when S is a type-II superconductor in the Meissner or diluted--vortex
phases. To be specific, we  consider exactly this case in Sec.~\ref{modelsec}, 
but discuss generalizations to the mixed phase in the Discussion section. The
motion in $z$ direction, i.e., parallel to the magnetic field, can be trivially
separated and gives only rise to an additive renormalization of the Fermi 
energy. We shall neglect this in the following, making our study especially
applicable for S--N junctions where the N region is a two--dimensional 
electron system. Eigenenergies for the motion in the $xy$ plane in a uniform
magnetic field are labelled by the guiding-center quantum number $X$ and 
correspond to the familiar Landau levels of independent electrons and holes
when $|X|\gg r_{\text{c}}$. Similar to the case of a hard
wall~\cite{bih:prb:82,pav:prb:84}, Landau levels are bent for $X$ close to the
S--N interface, and the corresponding eigenspinors of the BdG equation are
of mixed electron-hole type. An example of the spectrum of such solutions
with energies within the superconducting gap was given in Ref.~~\onlinecite
{uz:prl:00}.

The classical picture of skipping orbits along the interface, and the analogy
with the familiar quantum-Hall edge states,\cite{bih:prb:82,pav:prb:84,
butt:prb:88,haug:sst:93} suggests that currents will flow along the interface in
the plane perpendicular to $\vec B$  (i.e., in $y$ direction). In
superconductivity, {\em quasiparticle\/} current $\vec \jmath_{\text{P}} =\vec 
\jmath_{\text{e}}+\vec\jmath_{\text{h}}$ and {\em charge\/} current $\vec 
\jmath_{\text{Q}}=e(\vec\jmath_{\text{e}}-\vec\jmath_{\text{h}})$ are 
distinguished where, for a particular solution of the BdG equation, electron
and hole currents are given by\cite{degennes}
\begin{subequations}
\begin{eqnarray}
\vec\jmath_{\text{e}} &=& \frac{1}{m}\,\Re{\mathrm e}\left\{u^* \left(\vec p - 
e\vec A \right)u\right\}\quad ,\\
\vec\jmath_{\text{h}} &=& \frac{1}{m}\,\Re{\mathrm e}\left\{v \left(\vec p - 
e\vec A \right)v^*\right\}\quad .
\end{eqnarray}
\end{subequations}
It turns out~\cite{uz:prl:00} that $\vec\jmath_{\text{e}}$ and $\vec 
\jmath_{\text{h}}$ are always parallel to the interface for states with definite
guiding-center coordinate $X$ in the direction perpendicular to the interface.
This is no surprise because, owing to the peculiar quantum dynamics in a
magnetic field, the guiding center for electrons and holes cannot be localized
in $y$ direction in a state where it is fixed in the $x$ direction. Generalizing the
Hellmann-Feynman theorem\cite{hellfeyn} to the matrix BdG Hamiltonian of
Eq.~(\ref{bdg}), a general expression can be derived~\cite{uz:prl:00} for the
contribution of a particular eigenstate $(u_{n X},v_{n X})$ with energy
$E_{n X}$, labelled by Landau-level index $n$ and guiding-center quantum 
number $X$, to the {\em total\/} quasiparticle and charge currents in $y$
direction:
\begin{equation}
I_{n X}^{\text{(P,Q)}} = \int_{x} \left(\vec\jmath_{\text{P,Q}}\right)_y \quad .
\end{equation}
The total quasiparticle current turns out not to depend explicitly on the electron or
hole wavefunctions. Rather, it is given entirely in terms of the energy
spectrum,
\begin{equation}\label{qpcurrent}
I_{n X}^{\text{(P)}} = \frac{1}{\hbar}\,\frac{\ell^2}{L_y}\,
\frac{\partial E_{n X}}{\partial X}\quad ,
\end{equation}
which aquires a nonzero dispersion only close to the S--N interface. Hence,
confinement due to the interface with a superconductor gives rise to the
formation of a chiral {\em edge channel for quasiparticle current} in exactly
the same way as confinement due to a hard wall induces edge channels for
charge current~\cite{bih:prb:82,pav:prb:84}. We distinguish the new kind
of edge channel realized at the S--N interface from its counterpart near a hard
wall by calling it an {\em Andreev edge channel\/} and call the solutions $(u_{n
X}, v_{n X})$ of the BdG equation with nonzero $I_{n X}^{\text{(P)}}$ {\em
Andreev edge states\/}. [In Eq.~(\ref{qpcurrent}), $L_y$ denotes the system
size in $y$ direction.] As expected intuitively, the total charge current $I_{n
X}^{\text{(Q)}}$ flowing in an Andreev edge state {\em does\/} depend on
the details of the wave functions; it can be written as $I_{n X}^{\text{(Q)}}=I_{n
X}^{\text{(H)}}+I_{n X}^{\text{(S)}}$. The expression for $I_{n X}^{\text{(H)}}$ 
suggests a straightforward interpretation~\cite{uz:prl:00}. It is given by
\begin{equation}\label{hallcurrent}
I_{n X}^{\text{(H)}} = \frac{e}{\hbar} \,\frac{\ell^2}{L_y}\, \frac{\partial E_{n X}} 
{\partial X}\,\left(1 - 2\, B_{n X}\right)\quad ,
\end{equation}
where $B_{n X}=\int_{x} \left| v_{n X}\right|^2$ is the hole probability in the
Andreev edge state. 
Equation~(\ref{hallcurrent}) is the expression for current
in an edge channel that is {\em reduced\/} due to the presence of Andreev
reflection. For vanishing $B_{n X}$, i.e, when only normal reflection occurs at
the S--N interface, Andreev edge channels actually become the familiar
quantum-Hall edge channels~\cite{bih:prb:82,pav:prb:84,butt:prb:88,
haug:sst:93}. Inspection shows that the term $I_{n X}^{\text{(S)}}$ represents
the quasiparticle--conversion current in the superconductor that supports the
Andreev edge state. Note that the above discussion, as well as the analytic
expressions, for currents are generally valid, i.e., in particular also with finite
Zeeman splitting and diamagnetic screening currents present. It is then
possible~\cite{uz:prl:00} to express the Andreev--reflection contribution to the
S--N interface conductance in terms of the hole probabilties of zero--energy
Andreev edge states as
\begin{equation}\label{ARconduct}
G_{\text{AR}}=\frac{e^2}{\pi\hbar}{\sum_{n}}^\prime B_n, 
\end{equation}
where the label $n$ comprises the set of quantum numbers (Landau--level
index and spin--projection eigenvalue) for zero--energy eigenstates, and the
$\sum^\prime$ indicates that only electron-like states (i.e., those with $B_n<
1/2$) are to be summed over. As the experimentally relevant quantity, we present
results for $G_{\text{AR}}$ in the following.

We continue with presenting our model of an S--N junction subject to a
quantizing magnetic field in Sec.~\ref{modelsec}. (Readers not interested in
the mathematical formalism can skip this section.) Numerical results obtained
within that model are reported in Sec.~\ref{numresdisc}, where the effects of
Zeeman splitting and diamagnetic supercurrents are elucidated in detail. We
discuss implications of these results with particular emphasis on
experimentally realizable superconductor--semiconductor hybrid structures.
Conclusions are given in the final Sec.~\ref{conclude}.

\section{Model of an S--N hybrid system in a quantizing magnetic field
\label{modelsec}}

Our model of the S--N interface is similar, in spirit, to that used in the
 approach by Blonder, Tinkham, and Klapwijk~\cite{btk} (BTK) and
its recent generalizations to superconductor--semiconductor
interfaces.\cite{mort:prb:99} We consider a planar hybrid system (in the $xy$
plane) consisting of a semi-infinite two-dimensional electron system (2DES, 
located in the half-plane $x>0$) and a superconductor (occupying the
half-plane $x<0$). The effective mass of electrons, the Fermi energy, and the
modulus of the superconducting pair potential are assumed to be piecewise
constant as functions of the coordinate perpendicular to the interface:
\begin{subequations}\label{quantities}
\begin{eqnarray}
m(x) &=& m_{\text{s}}\,\,\Theta(-x) + m_{\text{n}}\,\,\Theta(x) \quad ,\\
\epsilon_{\text{F}}(x) &=& \epsilon_{\text{F,s}} \,\, \Theta(-x) + 
\epsilon_{\text{F,n}} \,\, \Theta(x)\quad ,\\
|\Delta|(x) &=& \Delta_0 \,\,\Theta(-x)\quad .
\end{eqnarray}
\end{subequations}
The magnetic field, applied perpendicular to the 2DES, is assumed to be
screened from the S region,
\begin{subequations}\label{meissner}
\begin{equation}
\vec B(x) = \left\{ B\,e^{\frac{x}{\lambda}}\,\,\Theta(-x) + B \,\, \Theta(x) \right\} 
\,\hat z\quad ,
\end{equation}
where $\lambda$ denotes the magnetic penetration depth. Translational
invariance in $y$ direction suggests a gauge for the vector potential where
$\vec A = A(x)\,\hat y$, and we choose
\begin{equation}
A(x) = \lambda \, B \, \left(e^{\frac{x}{\lambda}} - 1\right)\,\, \Theta(-x) + x \, B 
\, \,\Theta(x)\quad .
\end{equation}
\end{subequations}
In the presence of a magnetic field, the phase $2 \varphi$ of the 
superconducting pair potential aquires a non-trivial dependence on spatial
coordinates, and care has to be taken to determine it correctly~\cite{galaiko1,
svid:sjltp:75}. Combining Maxwell and London equations,
\begin{subequations}
\begin{eqnarray}
\vec\nabla\times\vec B &=&\mu_0 \,\vec\jmath_{\text{s}} \quad , \\
\vec\jmath_{\text{s}} &=& \frac{\hbar}{e\mu_0\lambda^2}\,\, \vec
k_{\text{s}} \quad ,
\end{eqnarray}
\end{subequations}
where $\vec\jmath_{\text{s}}$ is the screening supercurrent, and 
\begin{equation}
\vec k_{\text{s}} = \vec\nabla\varphi - e \vec A \quad ,
\end{equation}
we obtain from Eqs.~(\ref{meissner}) for the case under consideration
\begin{equation}
\varphi(y) = -\frac{\lambda}{l_B^2}\, y\,\,{\mathrm{sgn}}(e B) + \varphi_0\quad ,
\end{equation}
with $\mathrm{sgn}(x)$ denoting the sign function, and $\varphi_0$ a constant.

We calculate electronic and transport properties from solutions of the BdG
equation~(\ref{bdg}). Upon making the separation \textit{Ans\"atze\/}
\begin{subequations}\label{sepansatz}
\begin{eqnarray}
u(x,y) &=& \frac{1}{\sqrt{L_y}} \, f_{X}(x)\,\, e^{i\left[\varphi(y)+ y\, 
\frac{X}{l_B^2} \, {\mathrm{sgn}} (eB) \right]} \,\, ,\\
v(x,y) &=& \frac{1}{\sqrt{L_y}} \, g_{X}(x) \,\, e^{i\left[-\varphi(y) + y\, 
\frac{X}{l_B^2}\,{\mathrm{sgn}} (eB) \right]} \,\, ,
\end{eqnarray}
\end{subequations}
and using the fact that, in the presence of Zeeman splitting, Eq.~(\ref{bdg}) exhibits a block--diagonal form~\cite{kettsong} in the spin quantum number 
$\sigma$, we find a onedimensional BdG equation, 
\begin{equation}\label{1Dbdg}
\left(\begin{array}{cc} H_{+,\sigma} + U & |\Delta| \\ |\Delta| & 
- H_{-,-\sigma}-U \end{array}\right)\left(\begin{array}{c} f_{X,\sigma} \\
g_{X, -\sigma} \end{array} \right) = E \,\left(\begin{array}{c} f_{X,\sigma} \\
g_{X, -\sigma} \end{array}\right)\, .
\end{equation}
The single-particle Hamiltonians $H_{\pm,\sigma}$ are different in the S and
N regions, $H_{\pm,\sigma}=\Theta(-x)\,\, H_{\pm,\sigma}^{\text{(s)}} + 
\Theta(x) \,\, H_{\pm,\sigma}^{\text{(n)}}$, where
\begin{equation}\label{singpartN}
H_{\pm,\sigma}^{\text{(n)}} = \frac{\hbar\omega_{\text{c}}}{2} \left\{
\frac{l_B^2}{\hbar^2}\, p^2_x + \left[\frac{x + \lambda \mp X}{l_B}\right]^2 
+ \sigma \eta_{\text{Z}} - \nu \right\} \, .
\end{equation}
Here $\omega_{\text{c}}=\hbar/(m_{\text{n}}l_B^2)$ denotes the cyclotron
frequency in the N region, $\eta_{\text{Z}}=g\mu_{\text{B}} B/(\hbar 
\omega_{\text{c}})$ measures Zeeman splitting, and $\nu =2 \epsilon_{\text{F,n}}/(\hbar \omega_{\text{c}})$  is the filling factor. In the
superconductor, the single-particle Hamiltonian depends on the superflow 
wave vector $\vec k_{\text{s}}=k_{\text{s}}\,\hat y$,
\begin{subequations}
\begin{eqnarray}\label{supflow}
k_{\text{s}} &=& - \frac{\lambda}{l_B^2}\, e^{\frac{x}{\lambda}}\,{\mathrm{sgn}}
(eB) \,\,\Theta(-x) \quad , \\
H_{\pm,\sigma}^{\text{(s)}} &=& \frac{p^2_x}{2 m_{\text{s}}} + \frac{\hbar^2
\left[  X\pm l_B^2 k_{\text{s}}\,{\mathrm{sgn}}(eB) \right]^2}{2m_{\text{s}}
l_B^4}  - \epsilon_{\text{F,s}} \, .
\end{eqnarray}
Here we have neglected Zeeman splitting in the S region, which is a good
approximation for typical field strengths applied experimentally in S--2DES
hybrid structures. For our purposes, we need to find solutions of 
Eq.~(\ref{1Dbdg}) in the S region on length scales $x\sim\xi_0$ where $\xi_0$ 
is the superconducting coherence length. In a type-II material which has
$\xi_0\ll\lambda$, we can approximate $\exp\{x/\lambda\} \approx 1$ in
Eq.~(\ref{supflow}) and use
\begin{equation}\label{singpartS}
H_{\pm,\sigma}^{\text{(s)}} \approx \frac{p^2_x}{2 m_{\text{s}}} + \frac
{\hbar^2} {2 m_{\text{s}} l_B^4} \left( X\mp\lambda\right)^2 - 
\epsilon_{\text{F,s}} \quad .
\end{equation}
\end{subequations}
Our approximation amounts to neglecting the cyclotron motion of evanescent
quasiparticles in the S region, which translates into the condition that
the superconducting gap parameter $\Delta_0$ is much larger than the
cyclotron energy of electrons in the superconducting material. This is typically
satisfied for realistic situations.

Solutions of Eq.~(\ref{1Dbdg}) for the S--N system are found by matching
\textit{Ans\"atze\/} for eigenfunctions in the S and N regions at the interface
$x=0$. In the S region, where a uniform superflow with velocity $\hbar\lambda
/(m_{\text{s}}l_B^2)$ is present, the appropriate \textit{Ansatz\/} is a
superposition of Doppler-shifted quasiparticle excitations,
\begin{equation}\label{Sansatz}
\left(\begin{array}{c} f_{X,\sigma} \\ g_{X, -\sigma} \end{array}
\right)_{x<0}= d_- \left(\begin{array}{c} \gamma_- \\ 1
\end{array} \right) \, \psi_-(x) + d_+ \left(\begin{array}{c}
\gamma_+ \\ 1 \end{array} \right) \, \psi_+(x) \, ,
\end{equation}
with functions $\psi_\pm(x) = \exp\{\mp i x q_\pm\}$, and parameters
\begin{subequations}
\begin{eqnarray}
q_\pm &=& \left[ \frac{2 m_{\text{s}}}{\hbar^2} \left( \epsilon_{\text{F,s}} \pm \sqrt{(E+\delta_\lambda)^2 - \Delta_0^2}\right) - \frac{X^2+\lambda^2} {l_B^4}\right]^{\frac{1}{2}} , \\
\gamma_\pm &=& \frac{\Delta_0}{E+\delta_\lambda \mp \sqrt{
(E+\delta_\lambda)^2 - \Delta_0^2}}\quad , \\ \label{deltalambda}
\delta_\lambda &=& \hbar \omega_{\text{c}}\,\, \frac{m_{\text{n}}}{m_{\text{s}}}
\,\, \frac{X \lambda}{l_B^2} \quad .
\end{eqnarray}
\end{subequations}
For $|E+\delta_\lambda|<\Delta_0$, the wave function displayed in
Eq.~(\ref{Sansatz}) corresponds to evanescent quasiparticle excitations in S.
The \textit{Ansatz} for the wave function in the N region is a superposition of purely
electron and hole-like BdG spinors,
\begin{equation}\label{Nansatz}
\left(\begin{array}{c} f_{X,\sigma} \\ g_{X, -\sigma} \end{array}
\right)_{x>0}= \left( \begin{array}{c} a \\ 0 \end{array}\right)
\chi_{+,\sigma}(\zeta_+) + \left(\begin{array}{c} 0 \\ b\end{array}\right)
\chi_{-,-\sigma}(\zeta_-) \,\, ,
\end{equation}
where $\zeta_\pm = x + \lambda\mp X$, and the functions $\chi_{\pm,\sigma}
(\zeta)$ are eigenfunctions of the harmonic-oscillator Schr\"odinger equation,
\begin{equation}
\frac{l_B^2}{2}\,\frac{d^2\chi_{\pm,\sigma}}{d\zeta^2} -
\left[ \frac{\zeta^2}{2l_B^2} - \frac{\nu-\sigma\eta_{\text{Z}}}{2} \mp \frac{E}
{\hbar\omega_{\text{c}}} \right] \chi_{\pm,\sigma} = 0 \quad ,
\end{equation}
normalized to unity in the half-space occupied by the N region. They can be
expressed in terms of {\em  parabolic cylinder  functions\/}~\cite{abramowitz}
that are well-behaved as $x\to\infty$,
\begin{equation}
\chi_{\pm,\sigma}(\zeta_\pm) = F_{\pm,\sigma} \,\, U\left(-\frac{\nu-\sigma
\eta_{\text{Z}}}{2}\mp \frac{E}{\hbar\omega_{\text{c}}}, \sqrt{2}\,\,
\frac{\zeta_\pm}{l_B}\right)\, ,
\end{equation}
where $F_{\pm,\sigma}$ are normalization constants.
 
Imposing the continuity of BdG quasiparticle wave function and conservation
of the probability current at the interface yields the secular equation
\begin{widetext}
\begin{equation}\label{secular}
G H ({c^{\prime}}^2+{d^{\prime}}^2)+G^{\prime} H^{\prime}  =  c^{\prime}
\frac{E+\delta_\lambda}{\sqrt{\Delta_{0}^2-(E+\delta_\lambda)^2}}(G^{\prime} H-G H^{\prime}) + d^{\prime} (G^{\prime} H+G H^{\prime}) \quad , 
\end{equation}
\end{widetext}
whose parameters are:
$c^\prime=\frac{m_{\text{n}}}{m_{\text{s}}} \, \Re\text{e} \left\{ q_+ \right\}$, $d^\prime=\frac{m_{\text{n}}}{m_{\text{s}}}\, \Im\text{m} \left\{ q_+ \right\} +
\tilde U$ with $\tilde U=\frac{2 m_{\text{n}}}{\hbar^2} \, U_0$, 
$G=\chi_{+,\sigma}(\lambda-X)$, $H=\chi_{-,-\sigma}(\lambda+X)$, 
$G^\prime=\partial_x \chi_{+,\sigma}(x+\lambda-X)\left|_{x=0}\right.$, and 
$H^\prime=\partial_x \chi_{-,-\sigma}(x+\lambda+X)\left|_{x=0}\right.$.
Equation~(\ref{secular}) is an implicit relation between the guiding--center 
coordinate and the excitation energy $E$, yielding the dispersion relation $E$
vs.\ $X$. 

Since it is important for finding the Andreev--reflection contribution to the 
conductance, we also give the general expression for the amplitude ratio $a/
b$ of the wave function in the normal region:
\begin{equation}
\frac{b}{a}=\frac{G^\prime+G\left(\frac{E+\delta_\lambda}{\sqrt{\Delta_{0}^2-(E+\delta_\lambda)^2}} c^\prime-d^\prime\right)}{c^\prime H \frac{\Delta_0}{ \sqrt{\Delta_{0}^2-(E+\delta_\lambda)^2}}}.
\end{equation}

Further analytical progress can be made using the asymptotic
expansion~\cite{abramowitz} for parabolic cylinder functions $U(\epsilon,x)
\approx \frac{\Gamma(1/4-\epsilon/2)}{2^{\epsilon/2+1/4}\sqrt{\pi}}\cos 
\left[(\epsilon/2+1/4)\pi \sqrt{|\epsilon|} x\right]$, being $\Gamma$ the Gamma function. 
In our case this expansion is
valid when $\epsilon_{\text{F,n}}\pm(E-\sigma g\mu_{\text{B}} B/2) \gg
m_{\text{n}} \omega_{\text{c}}^2 (|X|+\lambda)^2$. With this, the 
approximated secular equation reads 
\begin{equation}\label{secularsemi}
\cos(\varphi_{+,\sigma}) + \Omega(\varphi_{-,\sigma}) = \frac{2 s}{s^2+w^2+1}\,
\frac{(E+\delta_\lambda)\,\sin(\varphi_{+.\sigma})}{\sqrt{\Delta_{\text{0}}^2 - (E+\delta_\lambda)^2}}\quad .
\end{equation}
Here we employed the Andreev approximation~\cite{andreev1,andreev2}, i.e.,
assumed $E, \Delta_{\text{0}}, |g\mu_{\text{B}} B| \ll \epsilon_{\text{F,n}},
\epsilon_{\text{F,s}}$, and used the abbreviations
\begin{subequations}
\begin{eqnarray}
\varphi_{+,\sigma} &=& \pi \left[\frac{E} {\hbar\omega_{\text{c}}} -\frac{\sigma
\eta_{\text{Z}}}{2}\right] + 2 \sqrt{\nu}\frac{ X}{l_{B}} - \frac{2}{\sqrt{\nu}} \left[ 
\frac{E}{\hbar \omega_{\text{c}}} -\frac{\sigma\eta_{\text{Z}}}{2}\right] \frac{\lambda}{l_{B}}
\quad , \\
\varphi_{-,\sigma} &=& \pi\,\frac{\nu}{2} +\frac{2}{\sqrt{\nu}} \left[\frac{E}  {\hbar\omega_{\text{c}}} -\frac{\sigma\eta_{\text{Z}}}{2}\right]\frac{X}
{l_{B}} -2 \sqrt{\nu}\frac{\lambda}{l_{B}}
\quad , \\
\Omega(\alpha) &=& \frac{[s^2+w^2-1]\sin(\alpha)+2 w \cos(\alpha)}
{s^2+w^2+1}\quad .
\end{eqnarray}
\end{subequations}
The parameter $s=[\epsilon_{\text{F,s}} m_{\text{n}}/(\epsilon_{\text{F,n}}
m_{\text{s}})]^{1/2}$ measures the Fermi-velocity mismatch for the junction,
and $w=[2m_{\text{n}} U_{\text{0}}^2/(\hbar^2 \epsilon_{\text{F,n}})]^{1/2}$
quantifies interface scattering. Within the same approximation, we can write
the amplitude  ratio as
\begin{widetext}
\begin{equation}
\frac{b}{a}=\mathcal{N} \frac{1-\sin(\varphi_{+,\sigma}+\varphi_{-,\sigma})-\left(
\frac{E+\delta_\lambda}{\sqrt{\Delta_{0}^2-(E+\delta_\lambda)^2}}s-w\right) 
\cos(\varphi_{+,\sigma}+\varphi_{-,\sigma})}{s \frac{\Delta_0}{\sqrt{\Delta_{0}^2-(E+\delta_\lambda)^2}}\left[\sin(\varphi_{+,\sigma})-\cos(\varphi_{-,\sigma})\right]}, 
\end{equation}
\end{widetext}
where
$$ \mathcal{N}=\frac{\Gamma[1/4(1 + \nu -\sigma\eta_{\text{Z}}) + E/
(2 \hbar\omega_{\text{c}})]}{\Gamma[1/4(1 + \nu +\sigma\eta_{\text{Z}}) - E/
(2\hbar\omega_{\text{c}})]}2^{E/(\hbar \omega_{\text{c}})} \quad . $$
In the following, we discuss the qualitative effects of finite Zeeman splitting
and screening supercurrents separately, based on our approximate analytic
approach described above. While this allows us to understand certain basic
features exhibited in the exact numerical results to be presented later, we
caution the reader that the quantitative estimates given in the
remainder of this section can be expected to apply only in limiting cases that
are consistent with our approximations.

\subsection{Finite Zeeman splitting, no screening current}

Here we discuss briefly the effect of the Zeeman splitting on Andreev edge 
states in the absence of screening currents. We consider the case of small
excitation energies, i.e. $|E-\sigma g \mu_{\text{B}} B| \ll \hbar \sqrt {\epsilon_{\text{F,n}}/(2 m_{\text{n}} X^2)}$. In this situation, we find
\begin{equation}\label{nonideal}
E_\sigma=\Delta_{\text{0}}\,\,\frac{(2n+1)\pi\pm{\text{arccos}}(
\Omega_{\text{0}}) - 2X\sqrt{2 m_{\text{n}}
\epsilon_{\text{F,n}}}/\hbar-\frac{\pi}{2} \sigma \eta_{\text{Z}}}
{q+\pi\Delta_{\text{0}}/
(\hbar\omega_{\text{c}})}\, ,
\end{equation}
where $\Omega_{\text{0}}=\Omega(\pi\nu/2)$ and $q=2s/(s^2+w^2+1)$.
Equation~(\ref{nonideal}) is very similar to the results obtained in the absence 
of Zeeman interaction in Ref.~~\onlinecite{uz:prl:00}. In particular, it shows
that, at low excitation energy (those that are relevant for linear transport), the 
Zeeman splitting can be absorbed in a spin--dependent shift of the
guiding--center coordinate $X$. We therefore expect the Zeeman splitting to
result in the usual halving of conductance steps as a function of filling factor.
As we will see below, however, this feature manifests itself only for very
transparent interfaces, while Zeeman splitting turns out to have a small effect
on the conductance in realistic situations.

\subsection{Effect of the screening current \label{screensec}}

It is immediately apparent from the general form of the secular equation that
there are two important consequences of finite screening currents. First, a
shift of quasiparticle energies in the superconductor by $\delta_\lambda$
occurs. [See Eq.~(\ref{deltalambda}).] As Andreev edge states have only
evanescent BdG spinor wave amplitudes in the superconductor, their 
formation is now possible only when the condition $\Delta_0>|E+ 
\delta_\lambda|$ is satisfied. Second, the spatial position corresponding to 
zero guiding--center coordinate of coupled electron and hole wave functions 
in the 2DES is shifted to $x=-\lambda$ into the superconducting region (see Fig. 
\ref{fig0}). Since the matching of BdG spinors still must be performed at the S--N interface ($x=0$), 
Andreev--reflection coupling in the normal regions becomes less direct.

For a more quantitative assessment of the effect of screening currents on the
Andreev--reflection conductance, we start by discussing the {\it ideal\/}
interface, i.e. $w=0$ and  $s=1$. In this case the secular equation reduces to
\begin{equation}\label{idscreen}
\cot(\varphi_+) =
\frac{E+\delta_\lambda}{\sqrt{\Delta_{\text{0}}^2 - (E+\delta_\lambda)^2}}
\quad .
\end{equation}  
For zero--energy bound states ($E=0$) and in the absence of Zeeman 
splitting, Eq.~(\ref{idscreen}) can be transformed into the secular equation for
the system \textit{without screening current\/} but at \textit{finite\/} energy
$\delta_\lambda$ and with a \textit{renormalized filling factor\/}
\begin{equation}
\nu^\prime = \nu\left(1 - \frac{\pi}{2}\,\frac{m_{\text{n}}}{m_{\text{s}}}\,
\frac{\lambda k_{\text{F,n}}}{\nu}\right)^2 \quad .
\end{equation}
This correspondence implies that sharp features in the linear conductance as 
a function of filling factor, which have their origin in the sudden appearance of
new bound--state levels, are shifted to larger filling factors due to the
presence of a finite screening current. In addition, the formation of
zero--energy Andreev edge states will be suppressed below a critical filling
factor
\begin{equation}
\nu_{\text{cr}}\approx \sqrt{\frac{m_{\text{n}}}{m_{\text{s}}}\,\frac
{\epsilon_{\text{F,n}}}{\Delta_0}\,\lambda k_{\text{F,n}}}
\end{equation}
because then their quasiparticle amplitudes in the superconducting region will
not be evanescent. Hence, the
Andreev--reflection contribution to the interface conductance vanishes for
$\nu<\nu_{\text{cr}}$.

In the more typical case of a \textit{nonideal\/} S--N interface, i.e., one with a
nonzero interface resistance, the screening current also renormalizes the
effective interface parameters $s$ and $w$. Considering again zero--energy
edge states in the absence of Zeeman splitting, we find new parameters
\begin{widetext}
\begin{subequations}
\begin{eqnarray}
w_\lambda &=& \frac{2 w \cos(2 k_{\text{F,n}}\lambda) - (s^2+w^2-1)\sin ( 2
k_{\text{F,n}}\lambda)}{s^2+w^2+1 -  (s^2+w^2-1)\cos(2 k_{\text{F,n}}\lambda)
-2 w \sin (2 k_{\text{F,n}}\lambda) } \, , \\
s_\lambda &=& \frac{2 s}{s^2+w^2+1 -  (s^2+w^2-1) \cos(2 
k_{\text{F,n}}\lambda) -2 w \sin (2 k_{\text{F,n}}\lambda) } \, .
\end{eqnarray}
\end{subequations}
\end{widetext}

\section{Numerical results and discussion\label{numresdisc}}

We have determined the BdG spinor wave functions for zero--energy Andreev
edge states numerically and calculated the Andreev--reflection contribution to
the S--2DES interface conductance according to Eq.~(\ref{ARconduct}). Here
we present results which elucidate the effects of finite Zeeman splitting and
superconducting screening currents.

\begin{figure}[b]
	\includegraphics[width=3.in]{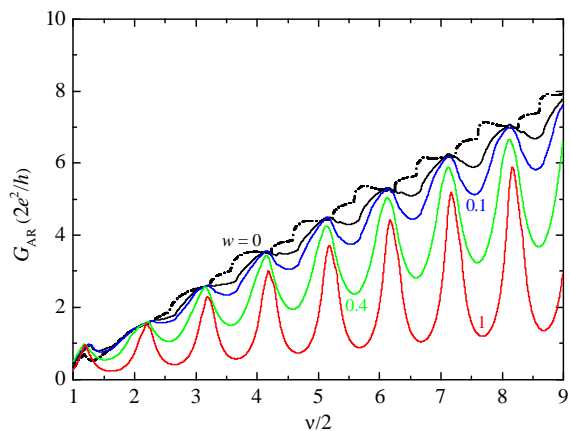}
\caption{Effect of Zeeman splitting on the Andreev--reflection contribution to
the S--N interface conductance $G_{\textrm{AR}}$. A double--step structure is clearly visible for
the dot-dashed curve which has been calculated for an ideal interface ($s=1$,
$w=0$) and $g=-20$, $\Delta_0=0.3$~meV, $\epsilon_{\text{F,s}} = 
\epsilon_{\text{F,n}} = 10$~meV, and $m_{\text{s}}=m_{\text{n}}=0.035 
m_{\text{e}}$. This feature turns out to be obscured at larger values of the 
superconducting gap ($\Delta_0=3$~meV for all solid curves), in particular 
when an interface barrier is present (nonzero values of $w$ as indicated for 
each curve).
\label{fig1}}
\end{figure}
\begin{figure}[b]
	\includegraphics[width=3.in]{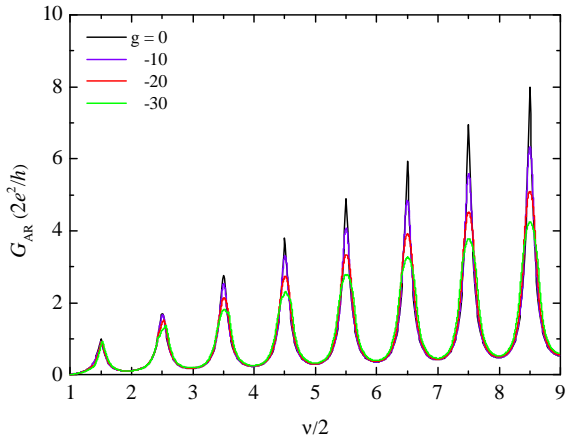}
\caption{Effect of Zeeman splitting for a realistic S--2DEG junction realized,
e.g., by a typical NbN/InGaAs hybrid system: $\epsilon_{\text{F,s}}= 7$~eV,
$\epsilon_{\text{F,n}} = 10$~meV, $m_{\text{s}}=m_{\text{e}}$, $m_{\text{n}} =0.035 m_{\text{e}}$, $\Delta_0 = 3$~meV, $w=0$, and for various values of
the gyromagnetic factor $g$ as indicated.
\label{fig2}}
\end{figure}

For the sake of clarity, we start by considering finite Zeeman splitting in
the absence of screening currents and for an ideal interface. As can be seen
from the dot--dashed curve in Fig.~\ref{fig1}, a double--step structure
emerges analogous to the conductance of spin--resolved
quantum--Hall edge channels~\cite{haug:sst:93}. This feature is quickly
suppressed, however, as the size of the superconducting gap increases, as
well as the interface barrier becomes more opaque($w>0$). Using parameters that
simulate a realistic NbN--InGaAs hybrid structure, Fig.~\ref{fig2} shows that the effect
of Zeeman splitting can be expected to be rather marginal in typical samples.
It results only in a suppression of the peak conductance, while the oscillatory
behavior as function of filling factor remains unaffected.

\begin{figure}[t]
	\includegraphics[width=3.in]{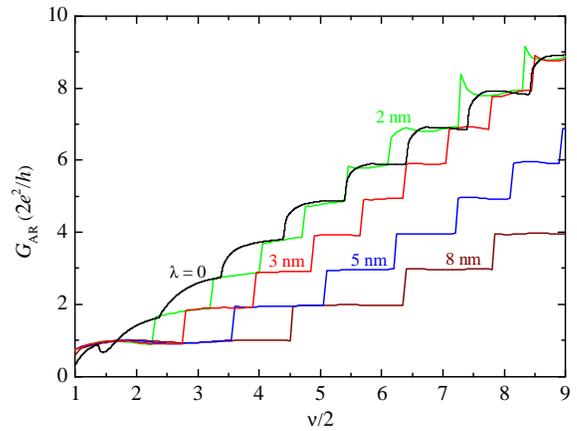}
\caption{Diamagnetic screening currents lead to a shift of conductance
steps. Data shown are calculated for an ideal interface ($s=1$,
$w=0$), $g=0$, $\Delta_0=0.3$~meV, $\epsilon_{\text{F,s}} = 
\epsilon_{\text{F,n}}  = 10$~meV, and $m_{\text{s}}=m_{\text{n}}=0.035 
m_{\text{e}}$ for various values of the magnetic penetration depth $\lambda$.
\label{fig3}}
\end{figure}
Turning to the investigation of the diamagnetic screening current, we show
numerical data for an ideal interface in Fig.~\ref{fig3}. As expected from our
analytical results presented in Sec.~\ref{screensec}, we find that step features
get shifted to higher filling factors as the magnetic penetration depth 
increases. The magnitude of the conductance is only marginally affected.
\begin{figure}[b]
	\includegraphics[width=3.in]{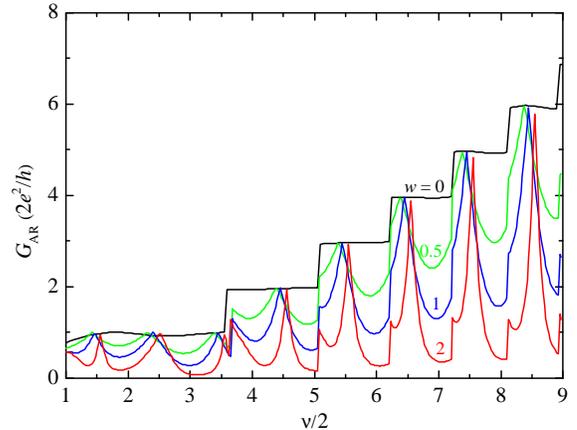}
\caption{Andreev--reflection transport at high magnetic fields through a
nonideal S--N interface with diamagnetic screening current present. Results
shown are for $\lambda=5$~nm and various values of $w$ as indicated, with
all other parameters as given in Fig.~\ref{fig3}.
\label{fig4}}
\end{figure}
Interestingly, an improvement of step quality is clearly apparent for increasing
penetration depth. As in the case without screening current, step--like features 
disappear and are replaced by an oscillatory conductance as soon as the
interface is nonideal. This can be seen in Fig.~\ref{fig4}, where the effect of 
interface scattering in the presence of screening currents is illustrated. While
some of the sharp features of the step structure remain, the Andreev--reflection
contribution to the conductance is much suppressed by an interface
barrier. It does reach a value comparable to the ideal case ($w=0$, $s=1$) 
only at specific values of filling factor.

\begin{figure}[b]
	\includegraphics[width=3.in]{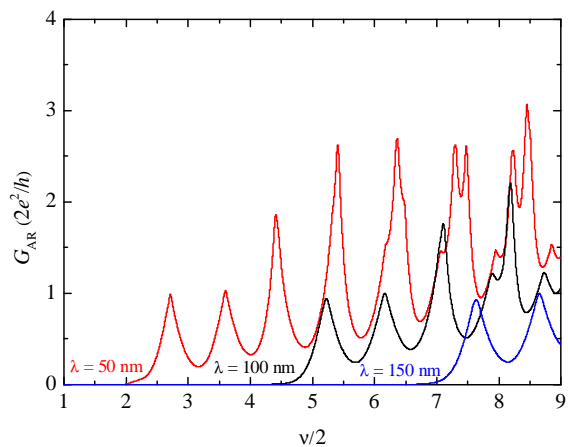}
\caption{Effect of the diamagnetic screening current at a realistic S--2DEG
interface such as in a NbN/InGaAs hybrid system, but for $g=0$. Sample
parameters are the same as for Fig.~\ref{fig2}.
\label{fig5}}
\end{figure}

Finally, the behavior of the conductance in a realistic S--2DES
hybrid structure with screening supercurrents present is shown in
Fig.~\ref{fig5}. Vanishing of the conductance below a critical filling factor
arises from the disappearance of zero--energy bound states due to the
Doppler shift of quasiparticles in the superconductor which reduces the
excitation gap to zero~\cite{degennes}. As the quasiparticles become
propagating in the S--region, Andreev reflection at the interface is 
suppressed.
This has important consequences for experimental investigation of Andreev
edge states. The results shown in Fig.~\ref{fig5} indicate that, for a S
electrode made from a NbN film for which $\lambda\sim 400$~nm has been 
reported in the literature~\cite{anlage}, effects due to Andreev reflection at the 
interface may be visible only at rather large values of the filling factor.

\section{Conclusions\label{conclude}}

We have investigated the interplay between Andreev reflection and cyclotron
motion at S--N interfaces. The effects of Zeeman splitting in the normal region
and diamagnetic screening currents in the superconductor have been taken
into account. In the ideal case, Zeeman splitting results in a doubling of step
features in the Andreev--reflection contribution to the interface conductance,
and screening currents lead to a shift of such features to larger filling factors.
However, in realistic S--2DES hybrid structures, the Zeeman splitting turns
out to merely suppress peak conductance. Screening currents turn out to be
important because the Doppler shift of quasiparticles in the S--region will
suppress their energy gap to zero below a critical filling factor, making
Andreev reflection disappear altogether. Above this critical value, the typical
oscillatory structure of the Andreev reflection contribution to the interface
conductance is displayed.

\begin{acknowledgments}

The authors wish to thank Rosario Fazio for fruitful discussions and for critical reading of the manuscript. UZ gratefully acknowledges the hospitality of Scuola Normale Superiore in
Pisa during a visit when work on this project was completed. This work was partially supported by FIRB "Nanotechnologies and Nanodevices for Information Society", contract RBNE01FSWY. 

\end{acknowledgments}

\end{document}